# Neutron β-decay, Standard Model and cosmology

A.P. Serebrov

*Petersburg Nuclear Physics Institute, Russian Academy of Sciences, Gatchina, Leningrad District, 188300, Russia*

A.P. Serebrov

Petersburg Nuclear Physics Institute

Gatchina, Leningrad district

188300 Russia

Telephone: +7 81371 46001

Fax: +7 81371 30072

E-mail: serebrov@pnpi.spb.ru

**Abstract**

The precise value of the neutron lifetime is of fundamental importance to particle physics and cosmology. The neutron lifetime recently obtained, $878.5\pm0.7_{stat}\pm0.3_{sys}$ s, is the most accurate one to date. The new result for the neutron lifetime differs from the world average value by $6.5\sigma$. The impact of the new result on testing of Standard Model and on data analysis for the primordial nucleosynthesis model is scrutinized.

## 1. Introduction, present status of neutron β-decay studies

The problem of precise measurements of the neutron lifetime is important for elementary particle physics and cosmology. The decay of a free neutron into a proton, an electron, and an antineutrino is determined by the weak interaction comprising the transition of a d-quark into a u-quark.

In the Standard Model of elementary particles, the quark mixing is described by the Cabibbo–Kobayashi–Maskawa (CKM) matrix which must be unitary. The values of the individual matrix elements are determined by the weak decays of the respective quarks. In particular, the matrix element $V_{\mathrm{ud}}$ can be determined from the data on nuclear β-decay and neutron β-decay. The extraction of $V_{\mathrm{ud}}$ from the data on neutron β-decay is extremely tempting due to the theoretical simplicity of describing the neutron decay compared to the description of nuclear decay. Unfortunately, the experimental procedure is a very complicated one, since it requires precise measurements of the neutron lifetime $\tau_{\mathrm{n}}$ and the β-decay asymmetry $A_0$.

The general formula for calculating $|V_{\mathrm{ud}}|^2$ is based on the neutron β-decay data $\tau_{\mathrm{n}}$ and $A_0$ [1]

$$|V_{\mathrm{ud}}|^2 = \frac{(4908.7 \pm 1.9)}{\tau_{\mathrm{n}}\left(1+3\lambda^2\right)} \ , \qquad (1)$$

$$f\tau_{\mathrm{n}}\left(1+\delta'_{\mathrm{R}}\right) = \frac{K}{|V_{\mathrm{ud}}|^2 G_{\mathrm{F}}^2 \left(1+3\lambda^2\right)\left(1+\Delta_{\mathrm{R}}\right)} \ , \qquad (2)$$

$$A_0 = 2\frac{\lambda(\lambda+1)}{1+3\lambda^2} \ , \qquad (3)$$

where $f$ is the phase space factor, $\delta_R'$ is a model-independent external radiative correction, $\Delta_R$ is a model-dependent internal radiative correction, $\lambda$ is the ratio of the axial-vector weak coupling constant to the vector coupling constant, $G_F$ is the Fermi weak coupling constant determined from the μ-decay, and $K = \hbar\ 2\pi^3 (\hbar c)^6 / (m_e c^2)^5$.

The formula (1) takes into account accuracy of calculation of radiative correction [1]. Thus, the required relative accuracy of measuring the neutron lifetime $\tau_n$ must be higher than $10^{-3}$ and than $2 \cdot 10^{-3}$ for $A_0$.

The accuracy of determination of neutron lifetime ($\tau_n$) and asymmetry ($A_0$) of neutron β-decay have been increased during the last 30 years about 20 times. Fig. 1 shows results of experimental data from 70s of the previous century up to the present time. The most precise measurement of $A_0$ has been reached by group of D. Dubbers from Heidelberg University. They announced a preliminary result: $A$=–0.1187(5) [2,3]. The corresponding $G_A/G_V$ -ratio is $\lambda$ = –1.2733(13). The most precise result for $A_0$ is in reasonable agreement with the value of PDG (2006) $A$ = –0.1173(13) and $\lambda$ = –1.2695(29).

The most precise measurement of neutron lifetime have been carried out by PNPI group with collaborators from JINR and ILL [4]. New experiment of neutron lifetime gives the value: 878.5±0.8 s, which differs from the PDG (2006) value (885.7±0.8 s) by 7.2 s or 6.5 standard deviations. In this experiment the method of storage of ultracold neutrons (UCN) in the trap was used, just as in the most previous experiments. However the probability of losses in experiment [4] was about 1% only of the probability of neutron β-decay. In previous experiments the loss factor was by 30 times higher. The extrapolation from the best storage time to neutron lifetime in new experiment was 5 s only, therefore it is very improbable to obtain systematic error about 7 s. The estimated systematic error in the new experiment was 0.3 s. Recently the measurement of neutron lifetime by means of continuous storage of UCN in magnetic trap has been carried out [5]. The result of measurement gives $\tau_n = \left(874.6^{+4.0}_{-1.6}\right)$ s. This value did not also confirm the world average value and is closer to the result of the experiment [4].

The next sections will be devoted to the analysis of the new result for neutron lifetime [4] and the most precise result of $A_0$ [2,3] for Standard Model and cosmology.

## 2. Standard Model with the most precision data of neutron β-decay

The main aim of precision measurements of neutron β-decay is to find deviations from the Standard Model assumptions as possible indications of new physics. $V_{ud}$ mixing matrix element obtained from neutron β-decay (${}^{n}V_{ud}$) can be compared with the matrix element obtained from $f\tau$ values for the nuclear superallowed $0^+ \to 0^+$ transitions (${}^{00}V_{ud}$). The superallowed $0^+ \to 0^+$ transitions are the pure Fermi transitions, and only $G_V$ coupling constant is involved in the process. The neutron decay proceeds through a mixed Fermi/Gamov-Teller transition and the both coupling constant $G_A$ and $G_V$ are involved in the process. In the case of pure *V-A* variant of interaction the both $V_{ud}$ values have to be equal to each other. Therefore the equation ${}^{n}V_{ud} = {}^{00}V_{ud}$ is the *V-A* test of Standard Model. Any other types of interaction can destroy this equation.

Another test of Standard Model is the unitarity test. $V_{ud}$ matrix element have to be equal to $\left(1 - V_{us}^2 - V_{ub}^2\right)^{1/2}$, where $V_{us}$ and $V_{ub}$ are determined from the transitions with s- and b-quarks. Thus, we have to check two equations:

$^{n}V_{ud} = {}^{00}V_{ud}$ (*V-A* test),

$^{n}V_{ud} = \left(1 - V_{us}^2 - V_{ub}^2\right)^{1/2}$ (unitarity test).

Fig. 2 shows analysis for both tests with data from PDG (2006). We can see very good agreement for $^{n}V_{ud}$, $^{00}V_{ud}$ and $\left(1 - V_{us}^2 - V_{ub}^2\right)^{1/2}$. Unfortunately the accuracy of λ-value from PDG (2006) is not so good to do precise conclusions for these tests. It would be more interesting to use the most precision last data.

Fig. 3 shows analysis for both tests with the most precision data of neutron β-decay. For this analysis the most precise data was used: the new value of $\tau_n$ = 878.5±0.8 [4] and the new value of $A_0$ = –0.1187(5) [2,3] or $\lambda$ = –1.2733(13). The inclined line demonstrates dependence of $^{n}V_{ud}$ from $\lambda$, which corresponds to equation (1). The vertical line is $\lambda$-value from [2, 3]. The crossing of these lines gives $^{n}V_{ud}$ value determined from neutron β-decay. We can compare $^{n}V_{ud}$ with $^{00}V_{ud}$ and with $\left(1 - V_{us}^2 - V_{ub}^2\right)^{1/2}$.

For *V-A* test we have the following result: $^{n}V_{ud} - {}^{00}V_{ud} = (2.4\pm1.0)\cdot10^{-3}$ or $2.4\sigma$.

As far as unitarity test is concerned the following equation can be written: $\left|{}^{n}V_{ud}\right|^2 + \left|V_{us}\right|^2 + \left|V_{ub}\right|^2 = 1.0038(28)$, $\Delta$= $-1.5\sigma$. In these calculation we use $V_{us}$=0.2257(21) from PDG (2006).

At last we can analyze the situation with new value of $A_0$ and value of $\tau_n$ from PDG (2006) 885.7±0.8 s. This dependence of $^{n}V_{ud}$ from λ is shown again in Fig. 3 by dotted line. In this case the difference $^{n}V_{ud} - {}^{00}V_{ud} = (-1.5\pm2.0)\cdot10^{-3}$ or $-1.5\sigma$ and the $\left|{}^{n}V_{ud}\right|^2 + \left|V_{us}\right|^2 + \left|V_{ub}\right|^2$=0.9981(28), $\Delta$= $+1.7\sigma$. The both differences do not exceed two standard deviations but have opposite sign with respect to previous analysis with new neutron lifetime data. It should be mentioned that averaging of old and new neutron lifetime data is impossible because the deviation is too big.

The deviation 2.4$\sigma$ mentioned above for ${}^{n}V_{ud} - {}^{00}V_{ud}$ with new value $\tau_n$ and with new value $A_0$ can not be considered yet as discrepancy. Therefore more precise measurements are required. The most important is the improvement of β-decay asymmetry measurements. However, the independent and precise measurements of neutron lifetime are needed as well as checking the result of the experiment [4] in spite of very high methodical level of new experiment.

**3. Neutron β-decay and cosmology**

Precise measurements of the neutron lifetime are also critically important when one wishes to verify a model of the early stages of the formation of the Universe. In the Big Bang model, at a temperature $T > 10^{10}$ K ($E >$ 1 MeV), the leptons, hadrons, and photons are in a state of thermodynamic equilibrium. As $T <$ 1 MeV, neutrinos are already incapable of sustaining this equilibrium state, since the rate of weak processes drops below the rate at which the Universe expands. The ratio $N_n / N_p$ of the number of neutrons to the number of protons at such a temperature is determined by the Boltzmann factor, $N_n / N_p = \exp(-\Delta m / T_f)$, where $\Delta m$ is the difference in the neutron and proton masses, and $T_f$ is the temperature at which weak-interaction reactions are frozen. Subsequently, the fraction of neutrons further decreases because of neutron β-decay. The nucleosynthesis process leads to the formation of deuterium and helium, mainly $^4He$. The abundance of these elements is determined by the ratio $N_n / N_p$. The key parameter that makes it possible to estimate the nucleosynthesis effect is the number of baryons per photon, $\eta_{10} = N_b / N_\gamma$. This parameter is related to the temperature and density of the early Universe and makes it possible to determine the conditions in which nuclear synthesis occurs. From this one can derive the initial element abundance. The following factors must be included in calculations of the nucleosynthesis of light elements: the rate $\Gamma_w = (7/60)\pi(1+3\lambda^2)G_F^2T^5$ of weak reactions, the rate of expansion of the Universe, the baryon asymmetry $\eta_{10}$, and the neutron lifetime $\tau_n$. Actually, the neutron lifetime enters the problem twice. It determines the rate of weak-interaction processes (the shorter the neutron lifetime, the faster weak reactions proceed and the earlier the neutrinos leave the state of thermodynamic equilibrium). In addition, a shorter neutron lifetime will produce fewer neutrons in the period when weak interactions become frozen and nuclear synthesis begins.

The observed quantities in the Big Bang model are the initial abundances of deuterium and $^4$He. These depend on the ratio of the number of baryons to the number of photons in the initial nucleosynthesis stage and on the neutron lifetime $\tau_n$. For instance, at a fixed value of $\eta_{10}$, a variation in the neutron lifetime by 1% changes the value of the initial abundance of $^4$He by 0.75%. The relative accuracy of measurement of the helium-4 abundance comprises ±0.61% [6]. Similarly, a variation in the neutron lifetime by 1% changes $\eta_{10}$ by 17%, although the modern accuracy of estimation of this quantity amounts to ±3.3% [6].

Detailed analysis of the nucleosynthesis process in the early stages of the formation of the Universe was recently made by Mathews et al. [6]. They analyzed the effect of the new value of the neutron lifetime on the consistency of data on the initial abundances of D and $^4$He isotopes and the data on baryon asymmetry $\eta_{10}$. Fig. 4 displays the dependences of the initial abundance of $^4$He ($Y_p$) on the baryon asymmetry $\eta_{10}$ taken from Ref. [6]. Clearly, the use of the new value of the neutron lifetime improves the agreement between the data on the initial abundances of deuterium and helium, and those on baryon asymmetry. Although the accuracy of the cosmological data is much lower than that of measurements of the neutron lifetime, the shift of $\tau_n$ from the world average value to the new value has a certain effect on the verification of the nucleosynthesis model in the early stages of the formation of the Universe.

**4. Conclusion**

The new measurement of neutron lifetime [4] improves the verification of the model of the early stages of the Universe formation as well as the verification of Standard Model. At the same time *V–A* test of Standard Model requires more precise measurements of *A*–asymmetry and independent precise measurements of neutron lifetime.

It should be mentioned that at present time there are several projects for precision measurements of $\tau_n$ and $A_0$ with relative accuracy $10^{-3}$ [3,7] therefore we can hope to obtain rather soon important information about tests of Standard Model from neutron β-decay.

**Figure captions**

Fig. 1. The progress in experimental accuracy of neutron β-decay data. a) *A*-asymmetry, b) neutron lifetime.

Fig. 2. Determination of $V_{ud}$ with PDG (2006) data: 1) from neutron β-decay data $\left({}^{n}V_{ud}\right)$, 2) from $0^+ \rightarrow 0^+$ nuclear transitions ${}^{00}V_{ud}$, 3) from unitarity. The inclined line is a dependence of ${}^{n}V_{ud}$ from $\lambda$ with PDG (2006) value of $\tau_n$=885.7±0.8 s. The vertical line is $\lambda$-value from PDG (2006): $\lambda$= –1.2695(29). The horizontal lines are $V_{ud}$ values from neutron β-decay, from $0^+ \rightarrow 0^+$ nuclear transitions and unitarity of CKM using $V_{us}$-value.

Fig. 3. Determination of $V_{ud}$ with the most precision data of $\tau_n$ and $A_0$: 1) from neutron β-decay data $\left({}^{n}V_{ud}\right)$, 2) from $0^+ \rightarrow 0^+$ nuclear transitions ${}^{00}V_{ud}$, 3) from unitarity.

Fig. 4. The predicted abundance of $^4$He – N($^4$He)/N(H) – in nucleosynthesis in the Big Bang as a function of the baryon-to-phonon ratio ($\eta_{10}$) with world average $\tau_n$ and with new neutron lifetime [4].

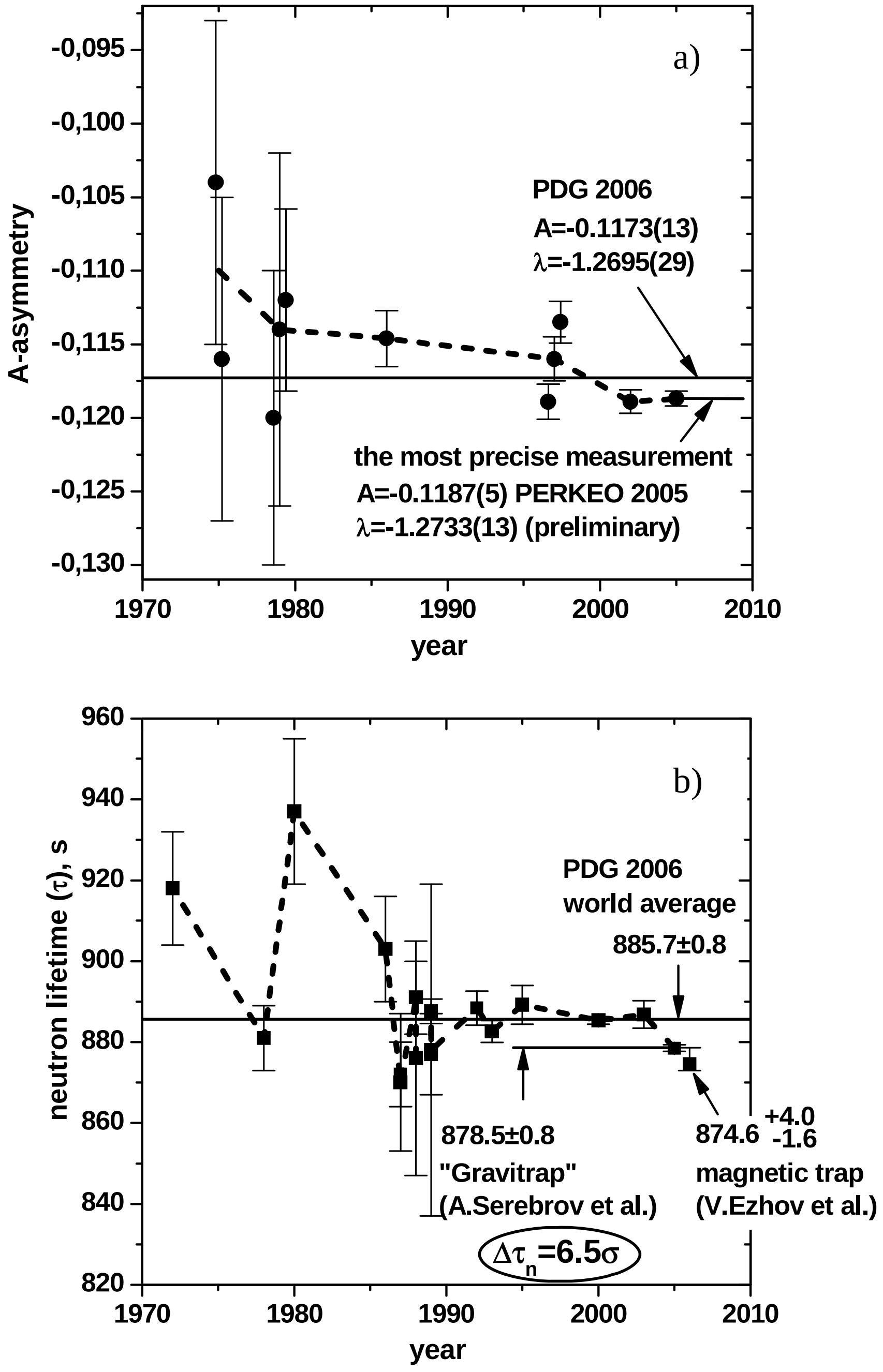

Fig. 1.

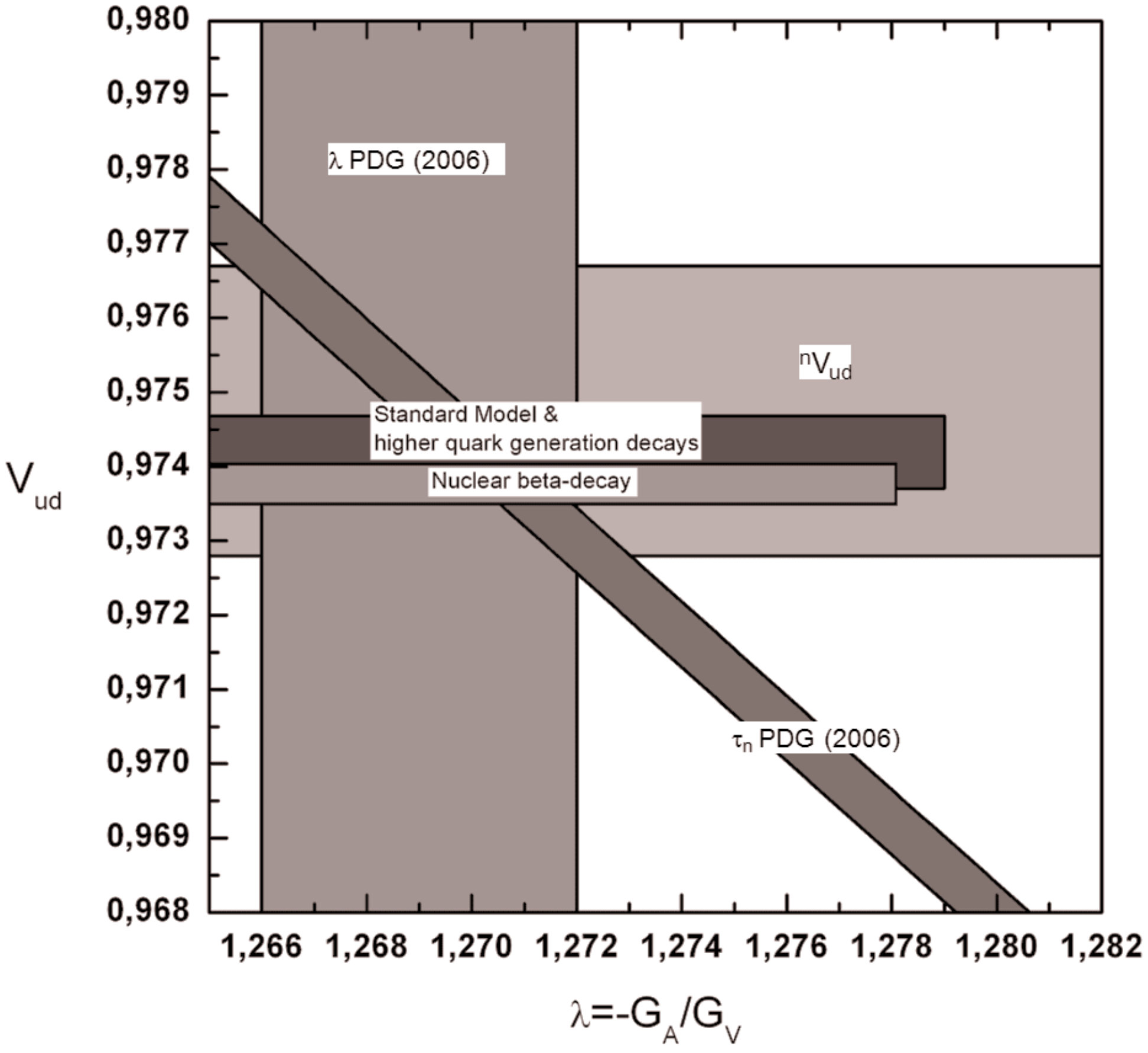

Fig. 2.

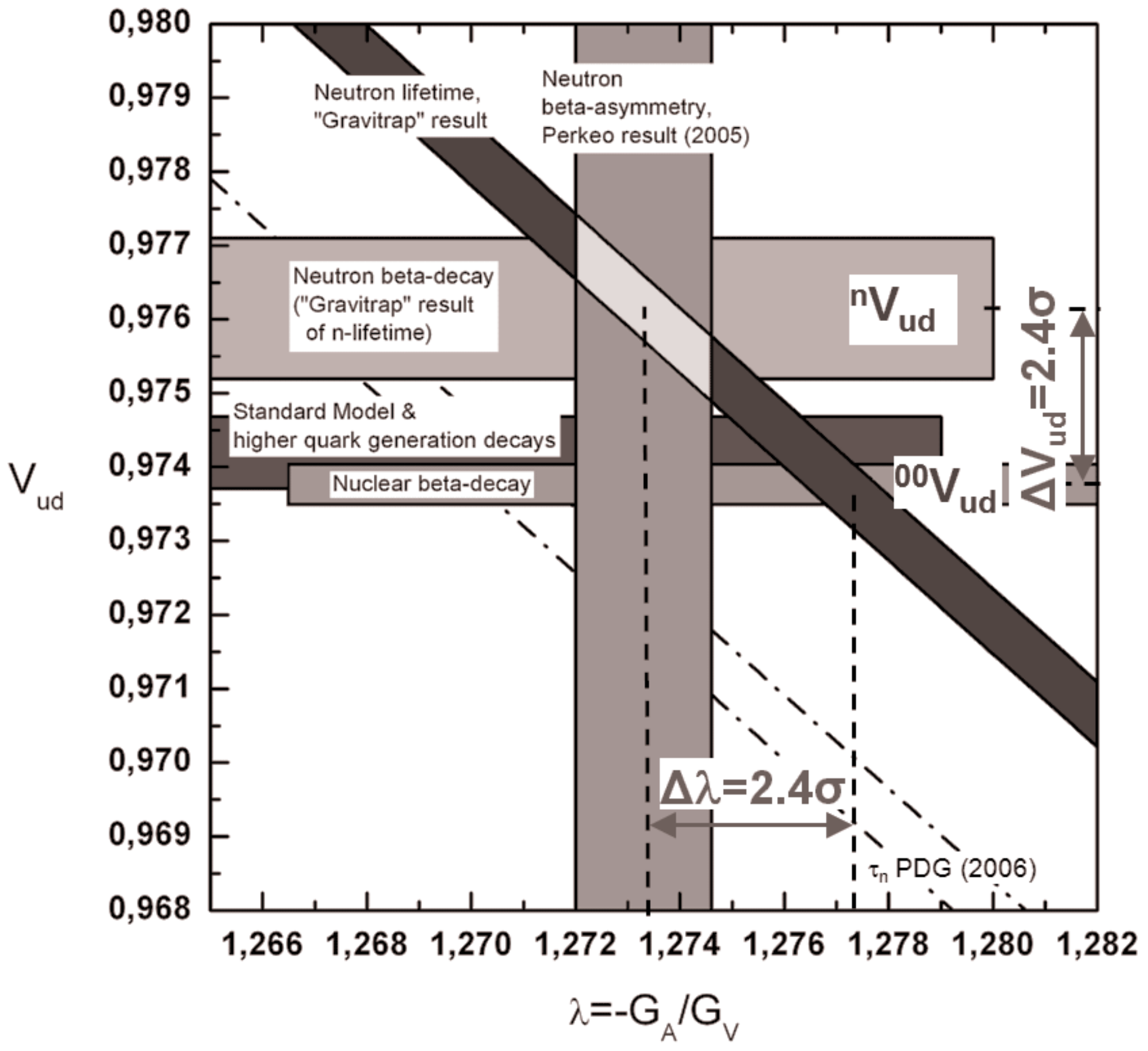

Fig. 3.

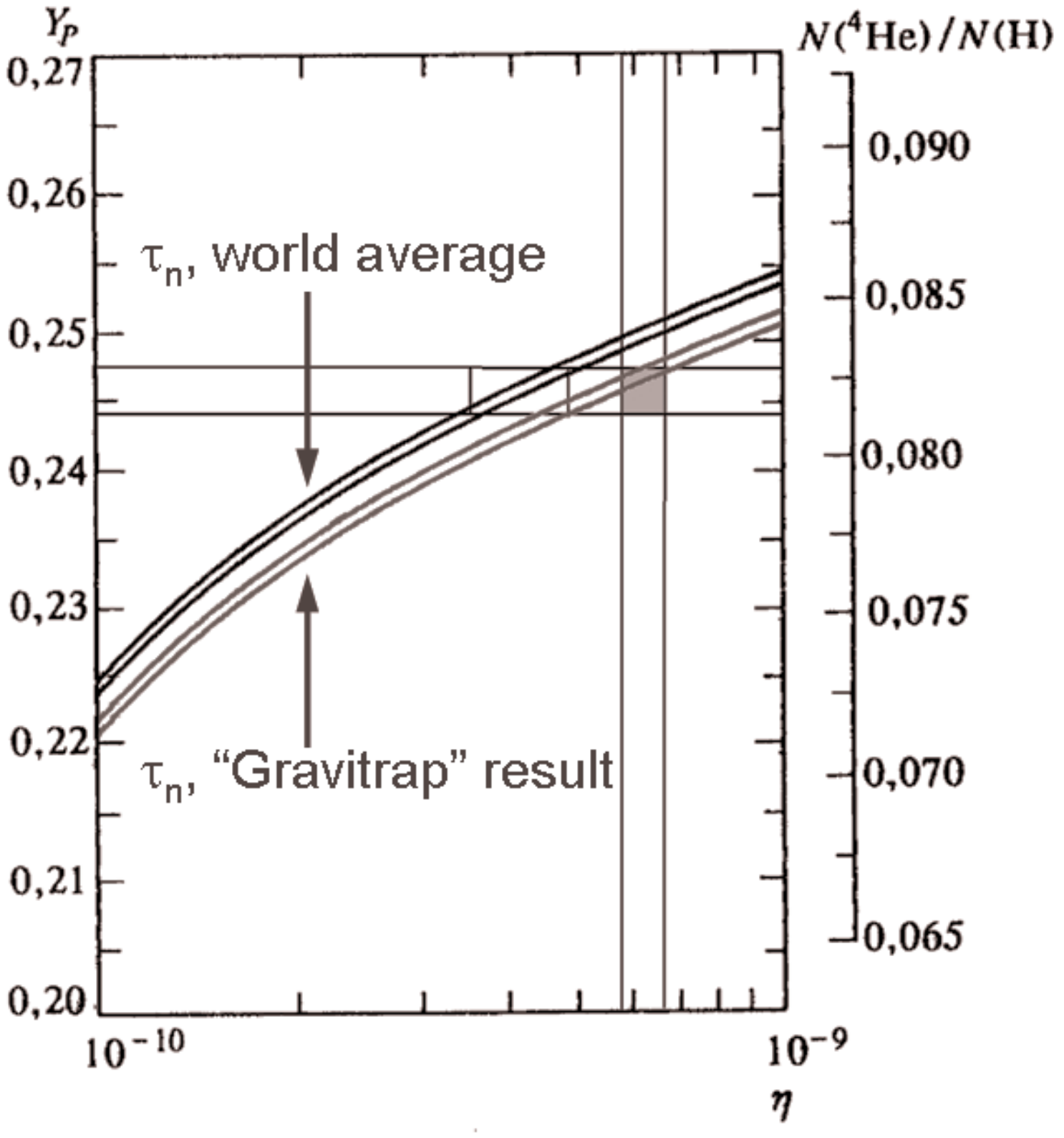

Fig. 4.